\documentclass[sigconf,nonacm]{acmart}
\usepackage{balance}
\setcopyright{none}
\renewcommand\footnotetextcopyrightpermission[1]{}
\title[Communication Shapes Collective Inference]{Communication Shapes
  Collective Inference in Self-Adapting LLM Societies: Evidence from
  Mafia}

\author{Haonan Huang}
\affiliation{\institution{Princeton University}\city{Princeton}\state{NJ}\country{USA}}
\author{Joey Xiao}
\affiliation{\institution{New York University}\city{New York}\state{NY}\country{USA}}

\begin{abstract}
When does communication help a group identify hidden adversaries, and
how does its value change as the group adapts? In Mafia, an informed
minority hides inside an uninformed majority whose only evidence is
open play. The zero-information game, where each day's vote eliminates
a random player, is exactly solved and scores every society;
matched-casting comparisons between protocols identify the effect of
communication. Societies of 8--100 \texttt{claude-haiku-4-5} agents (7{,}416
analyzed games, 1.9\,M model calls) adapt by rewriting and inheriting
private strategy notes. Simultaneous broadcast improves adversary
identification over silence in all nine compositions tested (8--46
players). Turn-taking removes most of this advantage; its voting
landslides are as frequent as broadcast's but land on mafia near
chance ($1.08\times$ versus $2.53\times$). At 70 players, agents
reading eight statements per day identify adversaries worse than
silent ones, and limited talk is worth less than at 46 players.
Adaptation is fast but need not help. In their first broadcast games,
citizens announce their role far more often than mafia (91\% vs.\
30\%) and first-day votes find mafia at three times chance; within two
generations citizens stop announcing and the cue fades, a change the
inherited notes carry. In controlled redeployments at 16 players,
societies carrying sixty generations of their own notes score below
societies with none. Communication shapes both collective inference
and the signals it depends on, so a protocol's value must be measured
together with the adaptation that changes those signals.
\end{abstract}

\keywords{LLM agents; multi-agent communication; social deduction; Mafia;
hidden-role games; exact baselines; cultural adaptation; collective
dynamics}

\newcommand{\fig}[2]{\IfFileExists{figs/#1}{\includegraphics[width=#2]{figs/#1}}{\fbox{\parbox{#2}{\centering figure \detokenize{#1} pending}}}}
\newcommand{\ci}[2]{[{#1},\,{#2}]}

\begin{document}
\maketitle

\section{Introduction}\label{sec:intro}

When does communication help a group identify hidden adversaries ---
and what does a society's own accumulated experience do to that
ability? Multi-agent LLM systems are increasingly built on the premise that
agents which talk~\cite{du2024debate,chateval2024} and agents which
remember~\cite{park2023generative,shinn2023reflexion} do better than
agents which do not; communication value is usually reported as an
improvement over a no-communication
ablation~\cite{lowe2019pitfalls,piatti2024govsim,ogara2023hoodwinked},
with neither arm on a common scale. And ``helps'' is one-sided: the
same channel can serve the hunted minority as well as the majority
hunting it. The quantitative question is what each way of organizing
communication does to the \emph{balance} between the two.

The hidden-role game Mafia supports a sharper design. An uninformed
majority (town) faces an informed minority (mafia) that eliminates one
citizen every night; each day the whole society eliminates one player by
plurality vote. The game's mathematics is unusually complete:
Braverman, Etesami and Mossel~\cite{braverman2008} showed that a fair
game needs a mafia of order the square root of the population
(revisited by Yao~\cite{yao2008}), and Migda{\l} \cite{migdal2010} solved the \emph{zero-information}
game (day votes blind, night kills unrestricted) in closed form, a
line that remains
active \cite{stock2026,tamashiro2026,knight2026}; the nearest analytic
treatment of LLM Mafia play is a capability model for a four-player,
single-day variant \cite{costa2026minimafia}. We use the
zero-information process as a fixed \emph{reference}: Section~\ref{sec:nullsim}
lays out its complete phenomenology, exactly and by simulation, and
every LLM society is measured as a per-elimination tilt $\theta$
against it --- while the value of communication itself is identified as
the \emph{paired difference between protocols on matched castings},
never as a departure from the reference, which non-linguistic channels
also move (Section~\ref{sec:null}).

On that ruler we place societies of \texttt{claude-haiku-4-5} agents at
$(n,m) \in \{(8,2), (12,3), (16,1..5),$ $(26,4), (46,5), (70,6),
(100,7)\}$, which adapt across generations by rewriting private strategy
notes after each game --- the only cross-game channel; all strategy
content is agent-authored. Communication is manipulated three ways: the
\emph{protocol} (simultaneous broadcast; sequential turn-taking;
silence; a private-planning turn), the \emph{bandwidth} $k$ (each player
reads a private uniform sample of $k$ statements per day), and an
informed \emph{detective}. Frozen evaluations redeploy fixed notes on
fresh castings, separating what was learned from the protocol it is
deployed into.

The paper reads the record in three layers, each with its own
standard of evidence: \emph{statics} (paired protocol contrasts on
matched castings), \emph{dynamics} (time-series tests against
surrogate nulls, and frozen redeployments that intervene on the
notes), and \emph{behavior} (associations in the game traces).

\textbf{Contributions.}
(1)~\emph{Measurement:} every lynch is scored against the exact
blind-chance rate (a logistic model with a known offset; a valid
partial likelihood because night kills remove only citizens), over
7{,}416 games whose records are re-hashed and audited; the reference
process is laid out exactly
(Sections~\ref{sec:null}--\ref{sec:harness}).
(2)~\emph{Statics:} broadcast talk shifts the balance toward the
majority, over silence, in 9 of 9 tested compositions; turn-taking
erases most of the shift; blank notes score above every evolved family in frozen broadcast
redeployment at one composition (resolved for five of six); talk at
eight or twelve statements per day is worth less at 70 players than at
46 (one family per arm), and full broadcast does not fit the context
window at 100 (Section~\ref{sec:statics}).
(3)~\emph{Dynamics:} the generation-1 tell and its collapse, shown by
frozen cross-deployment to travel with the inherited notes while
written advice mostly lags it (median first statement at generation
19, against a collapse at generation 2); and only broadcast societies show excess slow fluctuation and
family individuation, over outcomes with no detected trend
(Section~\ref{sec:dynamics}).
(4)~\emph{Behavior} (associations in the traces): lying is rare,
and mafia who claim citizenship are lynched more, not less; claims
mark citizens and draw the night kill; turn-taking herds statements,
not ballots, into cascades, and its landslides, as frequent as
broadcast's, land on mafia near chance --- about two-fifths of its
accuracy deficit (Section~\ref{sec:behavior}).

\section{Game Rules and Measurement}\label{sec:null}

\textbf{Pinned mechanics.} $n$ players, $m$ of them mafia; the mafia
know each other, citizens know only their own role. Each day, every
living player submits one public statement ($\le$80 words) and then
votes; the plurality choice is eliminated (ties uniform); we call the day
elimination a \emph{lynch} throughout. How statements are composed and delivered (all at once, one
speaker at a time, or not at all) is the manipulated \emph{protocol}
(Section~\ref{sec:harness}); voting
is always simultaneous, and the eliminated player's faction and the
full who-voted-whom record are published at the start of the next day.
Each night the mafia eliminate one citizen (rotating chooser). Town
wins by eliminating every mafioso; mafia win when they strictly
outnumber the remaining citizens. One elimination per phase is
mandatory, so game length is bounded.

\textbf{Measurement.} Write $n_t$ and $m_t$ for the living players and
living mafia when day $t$'s lynch is cast. A lynch blind to roles hits
mafia with probability exactly $m_t/n_t$ --- the living mafia's
population share. That blind process is the \emph{zero-information
reference}: its mafia win probability $w(n,m)$ is known in closed form
\cite{migdal2010} (e.g.\ $w(16,3)=0.576$, $w(100,7)=0.530$), and
Section~\ref{sec:nullsim} lays out how it plays. Against it, every
recorded lynch is one Bernoulli observation with
$\operatorname{logit} q_t = \operatorname{logit}(m_t/n_t) + \theta$:
the \emph{tilt} $\theta$ measures how far a society's lynches deviate
from blind chance --- by convention positive toward town (mafia found
faster), negative toward the informed minority. $\hat\theta$ maximizes
$\sum_t h_t\log q_t + (1{-}h_t)\log(1{-}q_t)$ over an arm's lynches;
because night kills remove only citizens, $(n_t,m_t)$ and the game's
end are functions of the lynch outcomes, so if the hit probability
given the past is $q_t$ this is the exact likelihood of the lynch
sequence; otherwise $\hat\theta$ is the tilt at which observed and
tilted-chance hits balance. All CIs are 95\% percentile bootstraps
resampling \emph{games} ($B{=}2000$, fixed seeds); contrasts between
arms sharing casting seeds are paired at the casting level.
``Resolved'' means the CI of a difference excludes 0; Section~\ref{sec:statics}
applies no multiplicity correction (the nine headline contrasts
survive Holm and Benjamini--Hochberg; supplement)\footnote{``Supplement'' refers to the supplementary material of the conference submission (verification detail, robustness tables, extended figures); it is omitted from this preprint and is available from the authors.}, Section~\ref{sec:dynamics}
corrects within test families; an additive-risk estimator agrees on
the headline contrasts (supplement). A generation-block bootstrap
leaves the headline contrasts unchanged and widens the scale intervals
of Section~\ref{sec:scale} by 17--32\% (supplement).

\textbf{Identification.} The reference is a fixed mechanical ruler, not
a claim about what any LLM society would do absent language:
\emph{silent} societies sit below it in six cells
(Section~\ref{sec:order}), consistent with the remaining channels moving elimination too: the
published ballot record, which lets an informed minority coordinate
ballots~\cite{wang2026werewolf}, and its choice of night victims
(the interventions that would separate the two were not run). No single-arm tilt is therefore read as ``the value of words'': the
estimand throughout is the paired protocol contrast on matched
castings. The game engine is verified against the exact theory (8
zero-information cells, 9 checks, max $|z|$ 1.43; a count-level chain
MC passes 52 more; supplement).

\section{The Zero-Information Game, Simulated}\label{sec:nullsim}

Before any language model plays, we lay out what the reference
society does. Each law below is computed exactly by dynamic
programming, agrees with Migda{\l}'s closed form \cite{migdal2010}
on all 1{,}767 even-$n$ states ($n\le120$), and is reproduced by
$2.1{\times}10^7$ independently coded Monte-Carlo games (max $|z|$
1.86 for win probabilities, 2.97 over all laws; supplement).

\textbf{The surface and its limit.} At fixed density
$x=m/\sqrt n$ the mafia win probability declines slowly with $n$ toward
a limit curve: taking $n\to\infty$ in the even-$n$ closed form of
\cite{migdal2010} gives $w(n,x\sqrt n)\to 2\Phi(x)-1$, with first
correction $+x^2\varphi(x)/\sqrt n$ (our term-by-term expansion;
derivation in the supplement; it differs from the parity-indexed
profiles \cite{tamashiro2026} proves under weak-majority stopping).
The dilute asymptotic \cite[Eq.~24]{migdal2010} is this limit's
linear term, and all corpus sizes collapse onto the curve (plotted in
the supplement).

\textbf{The fair line.} The $\sqrt n$ scaling of the fair mafia count
is classical \cite{braverman2008,yao2008}; the exact even-$n$ line
$m^*(n)$ --- where $w$ crosses $\tfrac12$ --- runs at
$m^*/\sqrt n = 0.646$--$0.657$ over our sizes ($\approx 0.65\sqrt n$),
and tends to $\Phi^{-1}(3/4)=0.6745$ as $0.6745-0.2275/\sqrt n$;
the dilute root $0.627\sqrt n$ \cite[Eq.~25]{migdal2010}, which
\cite{migdal2010} argues is exact for even $n$, understates it at
every $n$. The measured fair-line
shifts of Section~\ref{sec:order} are read against these exact values.

\textbf{How a null game unfolds.} With $\tau=2t/n$ the fraction of
the maximal span elapsed and $M$ the initial mafia count, the mean
mafia count tracks Migda{\l}'s law $m_t\approx M\sqrt{1-\tau}$
\cite[Eq.~21]{migdal2010} (within 0.2--1.5\% at mid-game) around a
widening fan (supplement), and the game is decided late: mafia wins are \emph{possible} only
in the last $m$ days, and at the five simulated sizes expected length is 91--97\% of the
$n/2$ bound
and 97--98\% of games end in those days. One structural fact holds for any
lynch rule, hence for every LLM game in this paper: with even $n$ the
night kill never ends the game, so \emph{every game ends at a day
lynch} --- the lynch sequence on which our partial likelihood conditions
carries every decision.

\begin{figure}[t]
\centering
\fig{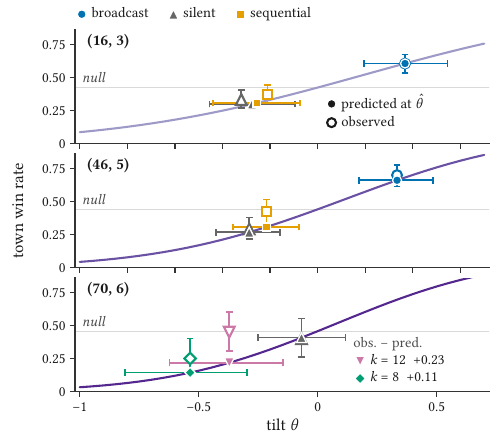}{\linewidth}
\caption{The tilt map: town win rate vs.\ per-lynch tilt $\theta$
at three sizes, with measured arms marked (filled: tilt prediction,
95\% CI; open: observed town rate, Wilson 95\%). The slope at
$\theta{=}0$ grows with $n$; the reference surface and trajectories
are in the supplement.}
\Description{Three stacked plots, one each for 16, 46 and 70 players, show how the town's chance of winning rises with the lynching tilt in the exact zero-information game, with each measured society marked on its curve and its observed win rate beside it; at 70 players the talking societies win more often than their tilt predicts.}
\label{fig:theory}
\end{figure}

\textbf{From per-lynch tilt to win rate.} The tilted process maps any $\theta$ to a town win rate
(Fig.~\ref{fig:theory}), and its
sensitivity grows with size at similar density:
$d\,\mathrm{town}/d\theta$ at $\theta{=}0$ is 0.49 at $(16,3)$, 0.66 at
$(46,5)$, 0.79 at $(100,7)$, because a fixed per-lynch edge compounds
over $\sim n/2$ lynches. Because the map's slope grows with $n$, a constant per-lynch edge
would buy more wins in larger societies; the decline reported in
Section~\ref{sec:scale} is in the per-lynch tilt itself, before any
mapping to wins.

\section{Societies, Adaptation, Harness}\label{sec:harness}

\textbf{Agents} (design schematic in the supplement). Every seat
runs \texttt{claude-haiku-4-5}\cite{anthropic2025haiku45}
(snapshot 20251001; extended thinking disabled, temperature 1.0,
decoding contract pinned and verified per request), living in its own session for the length of a game, whose growing
transcript is the agent's memory within that game: one plain reasoning-and-reply loop, no compaction or summarization,
driven by a Python orchestrator that pins and archives every request
at bare-API equivalence (detail in the supplement);
between games, the notes are the only channel.

\begin{figure}[t]
\centering
\fig{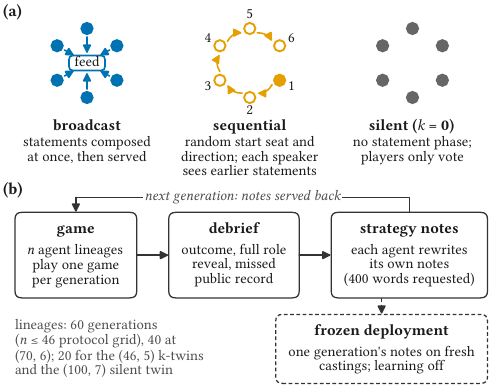}{\linewidth}
\caption{The manipulated protocols (top) and the adaptation loop
(bottom): after each game every agent rewrites its strategy notes,
served to the lineage's next game with roles re-randomized; frozen
evaluations redeploy fixed notes with learning off. Chains: 60
generations up to $(46,5)$, 40 at $(70,6)$, 20 for the twins.}
\Description{Design schematic: the three statement protocols, and the generational loop in which each agent is debriefed after a game and rewrites its strategy notes for the next generation, with a dashed branch where notes are frozen and redeployed.}
\label{fig:design-main}
\end{figure}
 One chain with a stronger mafia (\texttt{claude-sonnet-5}) in a
Haiku town is the only capability probe: town win rate down 0.22
\ci{0.02}{0.42} vs Haiku mafia, tilt still above the reference
($+0.43$ \ci{-0.01}{+0.81}); 30 games, not a model-dependence
test. Agents reply with one JSON object per turn and may keep a
private note ($\le$40 words); malformed replies draw bounded
corrections and can never stall the game --- a still-failing statement
simply publishes nothing (full fallback rules in the supplement).

\textbf{Protocols.} \emph{Broadcast}: statements composed simultaneously,
then served. \emph{Sequential}: one speaker at a time around a circle
with a uniformly random start seat and direction each day (not a uniform
random permutation); each speaker sees today's earlier statements.
\emph{Silent} ($k{=}0$): no statement phase; \emph{private
planning} replaces the statement slot by an unserved analysis turn
(used only in a control).

\textbf{Bandwidth.} At finite $k$, each player receives a private
uniform sample of $k$ of the other players' statements, in both
phases; draws come from a seeded per-(day, seat) stream and are
recorded.

\textbf{Adaptation and freezing.} Societies play repeated games in
lineages: after each game every agent sees the outcome, the full role
reveal and the public record it missed, then rewrites its strategy notes, a verbal-reflection loop
inherited across games \cite{shinn2023reflexion} (400 words
requested, not enforced; realized per-chain means 430--1{,}054),
served back next game with roles re-randomized. This is cultural transmission rather than
selection~\cite{perez2025telephone}: lineages accumulate and inherit
experience;
nothing is selected, scored or recombined (cf.\ LLM generations
\emph{with} selection of top performers~\cite{vallinder2025cultural}). A \emph{cell} is one
$(n,m)\times$protocol condition; each cell is run as two or three \emph{families} (A/B/C),
independent replicate societies with their own random castings (role
assignments and lineage-to-seat draws), and the protocols within a
family share identical castings, which is
what makes the paired contrasts of Section~\ref{sec:statics} possible.
Frozen evaluations redeploy one generation's notes on fresh castings
with learning off. Eliminated seats at finite $k$ see their missed days
at their own bandwidth (a paired rerun of one cell under the earlier
and later debrief contract finds no difference, $\Delta\theta=-0.02$
\ci{-0.31}{+0.29}; supplement).

\textbf{Corpus, audits.} The completed corpus holds 7{,}724
recorded games, 59{,}589
simulated game-days and 1{,}905{,}514 model calls; the analysis set of this paper is 7{,}416 games and 56{,}336
simulated game-days: the 6{,}778-game base battery (4{,}642 evolution
games in 81 chains, 2{,}136 frozen-deployment games in 49 cells, with
575k statements, 851k ballots and 94k lineage notes) plus the 638
confirmation games of Section~\ref{sec:scale}.
Integrity is checked end to end: record re-hashing (zero
mismatches), a 93.9\,GB byte-scan for tooling or identity leakage
(zero hits), and per-game contract audits (zero failures; supplement). One defect is
disclosed with its exact exposure: in a 15-day window, machine memory
pressure occasionally dropped a persisted turn from an agent's session;
549 games (11.8\% of the 4{,}642 analysis-set chain games; 0.12\% of
turns) were affected. Excluding them leaves broadcast--silent resolved in 9 of 9 cells;
two secondary contrasts lose resolution (supplement
table).

\begin{figure}[t]
\centering
\fig{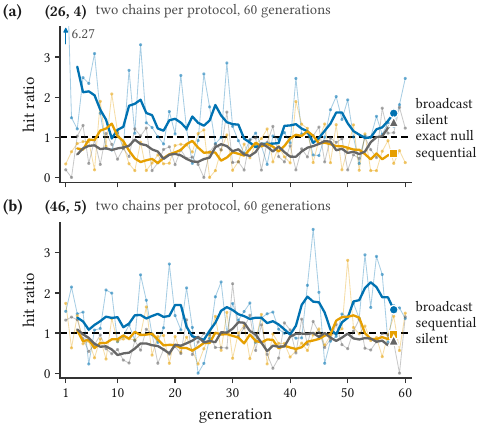}{\linewidth}
\caption{Hit ratio (hits/chance) by protocol at $(26,4)$ and
$(46,5)$: faint points pool the generation's games across families;
bold: centered 5-generation rolling mean. The late undulations are
what smoothing does to a persistent, aperiodic series; no periodicity
survives the surrogate null (Section~\ref{sec:dynamics}).}
\Description{Two line plots show, for societies of 26 and 46 players, how often lynches hit mafia members relative to chance in each of 60 generations, as faint per-generation points with bold five-generation averages for broadcast, sequential and silent societies around a dashed chance line.}
\label{fig:learning}
\end{figure}

\textbf{What a campaign looks like.} Figure~\ref{fig:learning} shows
two cells across their sixty generations: one game per family per
generation, the hit ratio (mafia lynched $\div$ blind-chance
expectation) wobbling around a protocol-specific level. Every statics number of the core grid pools exactly this record
(all generations and families of a cell, 120--180 games per point),
characterizing the whole history, not an endpoint;
Section~\ref{sec:dynamics} reads the same record in time.

\section{Statics: What the Protocol Does to the Level}\label{sec:statics}

Where does each society end up, relative to the exact reference?
This section carries the paper's headline numbers; every contrast is paired on matched castings (except $(26,4)$
sequential, compared unpaired on its own castings); the cross-size
comparison of Section~\ref{sec:scale} is a difference between two
such contrasts.

\subsection{Communication moves the balance; its form decides for whom}\label{sec:order}

\begin{figure*}[t]
\centering
\fig{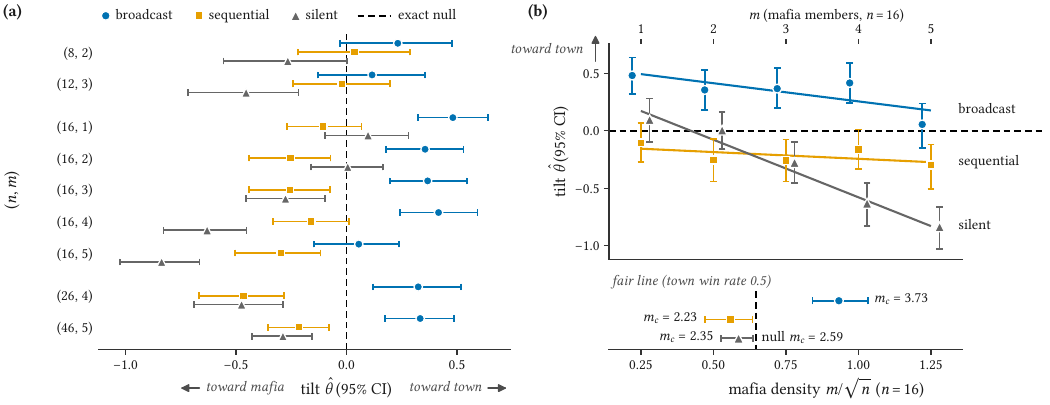}{\linewidth}
\caption{(a) Tilt $\hat\theta$ with 95\% game-cluster bootstrap CIs
for broadcast, sequential and silent societies at nine $(n,m)$ cells,
families pooled (120--180 games per point); dashed: the exact
reference; right of it favors town. (b) Density scan at $n{=}16$
(slopes per unit $m/\sqrt n$: broadcast $-0.32$, silence $-1.00$);
strip: measured fair lines (broadcast 3.73 \ci{3.37}{4.13} vs exact
2.59; silent 2.35; sequential 2.23).}
\Description{On the left, a forest plot gives the estimated lynching tilt with 95 percent intervals for broadcast, sequential and silent societies at nine compositions, every broadcast estimate lying right of the zero-information line, toward the town; on the right, the same estimates at 16 players against mafia density, with an arrow marking up as toward the town and a strip of measured fair lines.}
\label{fig:protocol}
\end{figure*}

Figure~\ref{fig:protocol} is the core comparison; in panel (a),
right of the reference line favors the town, left of it the informed
minority. Broadcast societies
sit toward town of their casting-matched silent societies in
\textbf{9 of 9} cells, $\Delta\theta$ from $+0.35$ at $(16,2)$ to
$+1.05$ at $(16,4)$; at $(46,5)$, $+0.62$ \ci{+0.44}{+0.81} over 120
paired castings, town rate 0.70 vs.\ 0.29. Sequential talk, with the same statement budget but speakers
hearing earlier statements before committing, recovers far less: it trails broadcast in 7 of 9 cells ($+0.35$ to $+0.79$; 6 of 9
when memory-affected games or the first ten generations are
dropped), exceeds silence in only 3, and sits \emph{below} silence at
$(16,2)$ ($-0.26$ \ci{-0.48}{-0.04}; not resolved once
memory-affected games are excluded). Silent societies never rise above the reference and sit below it
in 6 cells, the informed minority steering elimination away from
itself through the remaining channels, reaching
$\hat\theta=-0.84$ (town 0.028) at $(16,5)$.

Two facts sharpen the reading. First, \emph{the broadcast edge pre-dates adaptation}: in
generation 1, before any notes exist,
broadcast$-$silent is $+1.25$ \ci{+0.54}{+2.22} on 23 matched castings ---
also the only protocol comparison in the corpus with no training-history
difference at all (Section~\ref{sec:dynamics} identifies its
generation-1 carrier and what adaptation does to it). Second, \emph{density}: silence degrades at $-1.00$
\ci{-1.25}{-0.77} tilt per unit $m/\sqrt n$, broadcast at $-0.32$
\ci{-0.55}{-0.10}, so the gap grows at $+0.69$ per unit; sequential
is flat, crossing silence near $m{=}3$. In win-probability terms broadcast lifts the measured fair line from
the exact 2.59 to 3.73 \ci{3.37}{4.13}; silence and sequential fall
below it (Fig.~\ref{fig:protocol}b).

Excluding the 549 memory-affected games (Section~\ref{sec:harness})
leaves broadcast--silent resolved in 9/9 and changes three secondary
contrasts; pooling hides real family structure (per-family tables and
the robustness detail in the supplement).

\subsection{Bandwidth: response shapes, not yet a law}\label{sec:bandwidth}

Redeploying one family's fixed notes across $k$ asks how much of
the channel the deployed value needs: where the response crosses the
reference it does so at $k_c\approx5$--$7$ statements per day at the
mid sizes (roughly $1.3m$), but three of six sizes never cross at all,
and a parametric saturation fit cannot tell constant, $\sqrt n$,
linear-$n$ or $\propto m$ scalings apart (curves, per-size deployment
effects and fits in the supplement). We make no $k_c(n)$-law
claim.

\subsection{Transfer: family-idiosyncratic, and worse than no notes}\label{sec:transfer}

The $2{\times}2$ cross-over at $(16,3)$
initially suggested a striking rule: family
A's broadcast-trained notes did better silenced, and its silent-trained
notes better broadcast (interaction $-0.81$ \ci{-1.32}{-0.25}). The
pre-registered replication with two further note-writing families
refutes the rule while confirming the instability: family B shows
the same sign without resolving, family C the \emph{opposite} sign,
and the pooled interaction is null ($-0.19$ \ci{-0.54}{+0.17};
figure in the supplement). What evolved notes do under redeployment depends on the family that
wrote them, not systematically on the protocol they grew under.

(The matrix's blank-notes baseline is reported with the dynamics
evidence, Section~\ref{sec:dynamics}.) The same-notes broadcast-vs-sequential contrast resolves only for
blank notes ($+1.14$ \ci{+0.53}{+1.82}): the protocol effect needs no
adaptation, and the erasure of Section~\ref{sec:order} stands as a
result about co-adapted societies.

Two bounded controls (detail in the supplement): an
audience-free \emph{private-planning} turn did not lift the deployed
notes in either family tested ($-0.22$, $-0.09$; compute not matched);
and societies \emph{evolved at reduced bandwidth} ($k{=}4$) land at
their silent twins' level ($+0.10$ \ci{-0.20}{+0.40}), $-0.43$
\ci{-0.73}{-0.15} below their full-bandwidth twins --- a result
Section~\ref{sec:scale} must reckon with.

\subsection{Scale: no value of limited talk at (70,6); full
broadcast does not fit at 100}\label{sec:scale}

\begin{figure*}[t]
\centering
\fig{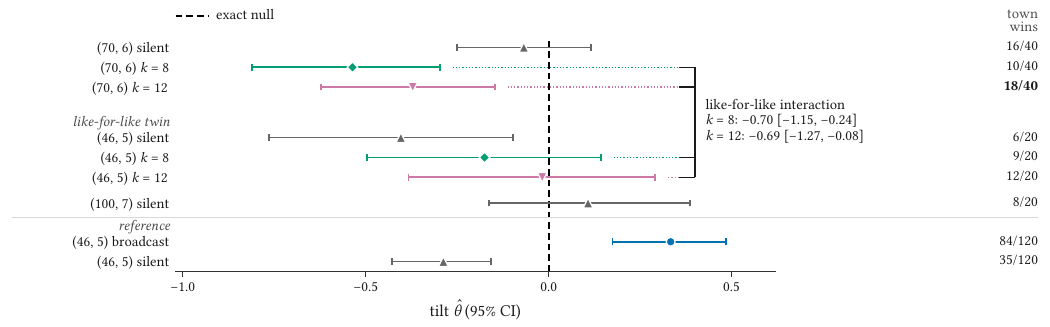}{\linewidth}
\caption{$\hat\theta$ at large $n$: the three matched $(70,6)$ arms
(40 generations each), the $(100,7)$ silent twin, the $(46,5)$
full-corpus references, and the like-for-like twin (silent
generations 1--20; evolved $k{=}8$, $k{=}12$); the bracket marks the
cross-size interactions; town win counts at right. The exploratory
by-phase decomposition is a supplement figure.}
\Description{A forest plot of lynching tilt for the 70-player arms,
their 46-player twins and references, with a cross-size interaction
bracket; and an exploratory panel showing the talk-minus-silent
deficit concentrated in the early game at 70 players.}
\label{fig:amp}
\end{figure*}

At $(70,6)$ --- three 40-generation chains with matched castings:
silent, $k{=}8$, $k{=}12$; full bandwidth does not fit ($k{=}8$ seats peaked at 188.5k of
192k usable tokens; supplement) --- no arm rises above the reference, and talk at
$k{=}8$ sits resolvably \emph{below} silence: silent $-0.07$
\ci{-0.25}{+0.12} (town 16/40), $k{=}8$ $\mathbf{-0.54}$
\ci{-0.81}{-0.30} (10/40), $k{=}12$ $-0.37$ \ci{-0.62}{-0.15}
(18/40); paired $k8{-}k0=-0.47$ \ci{-0.71}{-0.23} (resolved),
$k12{-}k0=-0.30$ \ci{-0.64}{+0.01} (not; its town count is higher,
18 vs 16).

Evolving at reduced bandwidth can fail for other reasons
(Section~\ref{sec:transfer}), so the cell alone confounds scale with
limited-bandwidth adaptation. We pre-registered the arbitration: an exact one-size-down twin at
$(46,5)$, $k{=}8$/$12$, 20 generations, castings paired to the silent
chain. At $(46,5)$ both limited-bandwidth arms sit above silence in point
estimate ($k8{-}\mathrm{silent}$ $+0.23$ \ci{-0.16}{+0.61},
$k12{-}\mathrm{silent}$ $+0.39$ \ci{-0.14}{+0.88}, $\Delta$town
$+0.30$ \ci{0.00}{+0.60}; full bandwidth on the same castings
$+0.80$ \ci{+0.30}{+1.29}); by the pre-registered rule (either CI above 0 or point estimate
$>{+}0.15$) this selects the size reading, though neither contrast
resolves alone. With all 40 $(70,6)$ generations the like-for-like cross-size
interaction --- evolved arms on both sides, matched designs --- is
resolved at both bandwidths:
$[k{-}\mathrm{silent}]_{(70,6)}-[k{-}\mathrm{silent}]_{(46,5)}=-0.70$
\ci{-1.15}{-0.24} at $k{=}8$ and $-0.69$ \ci{-1.27}{-0.08} at
$k{=}12$; on matched 20-generation windows it is $-0.61$
\ci{-1.19}{-0.07} at $k{=}8$ and $-0.54$ \ci{-1.28}{+0.16} at
$k{=}12$ (unresolved). Under a circular generation-block bootstrap the $k{=}8$ interaction
and $k8{-}k0$ hold at every block length, while the $k{=}12$
interaction is borderline at block length 5 (supplement). Talk at
these bandwidths is worth less at 70 players than at 46. The $(46,5)$ silent comparator has 7 of 20 generations in the
memory-defect window; dropping them strengthens both interactions
($-0.85$, $-0.93$), a conservative exposure. One measured misfit is reported with it: at $(70,6)$ the scalar
tilt under-predicts the talking arms' town wins (observed minus
tilt-predicted town rate $+0.23$ \ci{+0.15}{+0.30} at $k{=}12$,
$+0.11$ \ci{+0.02}{+0.18} at $k{=}8$, $-0.01$ silent;
Fig.~\ref{fig:theory} marks every such gap, including the $(46,5)$
sequential arm's $+0.12$). Two exploratory readings address this and the long-transcript
objection: the talking arms' deficit is front-loaded (days 1--10:
$-0.75$ at $k{=}8$, $-1.27$ at $k{=}12$; day 21 on: $-0.30$ \ci{-0.67}{+0.04} and $+0.13$
\ci{-0.32}{+0.59}; supplement figure), when transcripts are
short; on day-matched windows (days 11--20) the interaction is resolved
at both bandwidths ($-1.11$, $-1.36$); and a bulk/endgame refit
resolves the $k{=}12$ misfit
($\theta_{\mathrm{end}}-\theta_{\mathrm{bulk}}=+1.33$; predicted
town 0.50 vs observed 0.45) but not $k{=}8$. Comparability: each arm is one
family; the $(46,5)$ silent comparator predates its $k$-arms (served
texts byte-equal under masking), while the $(70,6)$ trio ran
concurrently.

The $(100,7)$ silent twin plays at the reference ($+0.11$
\ci{-0.16}{+0.39}), declining late ($-0.89$ \ci{-1.59}{-0.30}); two
single finite-$k$ generations are descriptive only. Full-bandwidth talk at $n{=}100$, however, is not merely expensive;
it does not fit this architecture: a measured $k{=}20$ game overflowed a surviving seat's
200k-token context by day 38, and at the measured per-day token costs a
full-bandwidth survivor would need $\sim$330k (measurement, proactive
stop and projection separated in the supplement). The bound is a property of this model, memory and protocol stack
(summarization, output caps or longer contexts would move it), but
within the stack it is a hard trade: open discussion's context cost
grows with lifetime $\times$ bandwidth (cf.\ the
scale--expressiveness trade-off of LLM
ABMs~\cite{chopra2025limits}), and the measured value at the
bandwidths that fit had already vanished.

A detective variant at $(26,4)$, in which one citizen privately
learns one true role per night, did not resolvably raise the town's
win rate (0.650 \ci{0.56}{0.73} vs.\ 0.608 for matched twins; tilt
lift $+0.16$ \ci{-0.12}{+0.45}, family A $+0.63$, family B $-0.32$)
and stays below an exact benchmark that adds a trusted detective
to the twins' measured edge (0.770; binomial $p{=}0.003$; conventions
in the supplement).
Section~\ref{sec:behavior} shows one reason disclosure is costly.

The bandwidth, detective, density, cross-over and scale tests were
pre-registered: hypotheses and decision rules fixed before the data,
and the supplement lists each with its outcome, including the failed
and refuted predictions. The dynamics and behavior analyses of
Sections~\ref{sec:dynamics}--\ref{sec:behavior} are exploratory
analyses of the locked records.

\begin{figure*}[t]
\centering
\fig{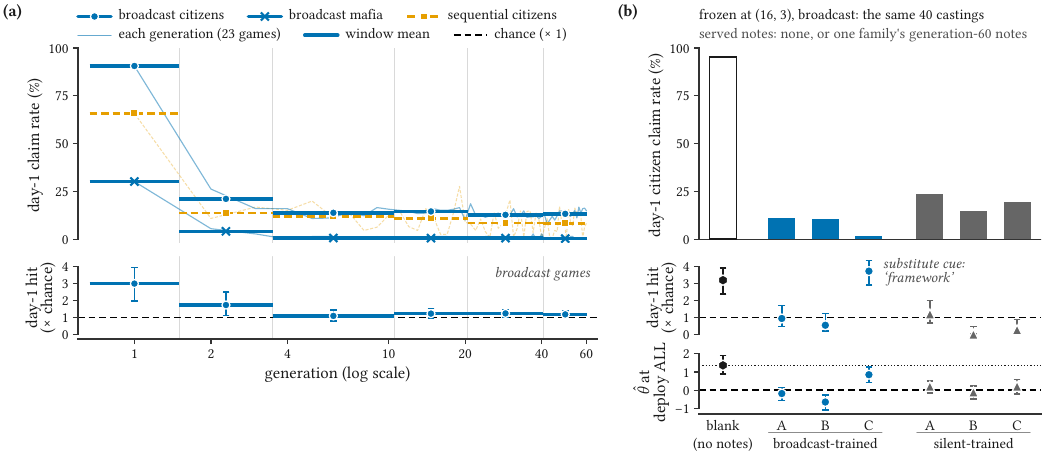}{\linewidth}
\caption{The tell and its self-destruction. (a) First-day role-claim
rates in broadcast societies by generation (window means bold;
sequential citizens dashed; log axis), with the first-day mafia-hit
factor per window below: the $3\times$ edge decays as citizens stop
claiming; mafia claims never return. (b) Frozen redeployment on 40
identical castings at $(16,3)$: every generation-60 note set
suppresses the claims; family C's broadcast-trained notes install a
substitute cue (hits $3.73\times$). Bottom strip: $\hat\theta$ at
full deployment; evolved $-$ blank resolved for five of six
families.}
\Description{On the left, line plots over 60 generations show the share of day-1 statements in which broadcast citizens, broadcast mafia and sequential citizens say they are citizens, with the day-1 rate of lynching mafia relative to chance below, both dropping sharply after the first generation; on the right, for frozen 16-player games on the same 40 castings, bars and strips compare blank notes with six evolved note sets on claim rate, day-1 accuracy and overall tilt, blank notes giving the highest claim rate and tilt.}
\label{fig:dynamics}
\end{figure*}

\begin{figure}[tb]
\centering
\fig{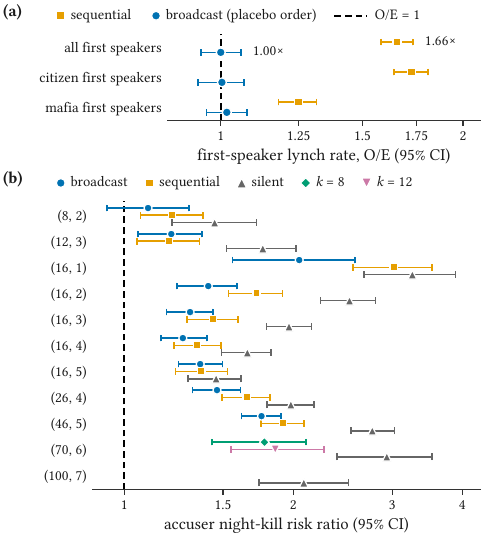}{\linewidth}
\caption{(a) First-speaker lynch rate over the position-blind rate:
sequential 1.66 \ci{1.58}{1.73} vs.\ the broadcast placebo (the engine's
own order draw) 1.00; within role, citizens 1.72, mafia 1.25.
(b) Ballot-accuser night-kill risk ratio --- citizens whose ballot went
to a mafia member that day --- one point per arm (log scale, reference
1): 1.10--3.26, CI above 1 in 30 of 31 arms, highest under silence in
9 of 9 cells. Associations in the recorded games; no causal claim.}
\Description{Two forest plots on logarithmic axes: how often the day's first speaker is lynched relative to a position-blind expectation under sequential play and in a broadcast placebo, overall and by role; and the night-kill risk ratio of citizens whose ballot targeted a mafia member, for every arm.}
\label{fig:mech}
\end{figure}

\section{Dynamics: From the Prior to the Adapted Society}\label{sec:dynamics}

How do societies travel from the model's untrained prior to adapted
play? Claim statistics use extractors validated blind (supplement on
validation provenance); time-series tests use permutation or
AR(1)-surrogate nulls with Benjamini--Hochberg correction within each
family of tests.

\textbf{The prior and its tell.} Untrained societies arrive with an
honest convention. In the first games of the 23 casting-matched
broadcast chains, citizens open with an explicit ``I am a citizen''
(91\% of 351 statements) and mafia mostly do not (30\%), so a missing
claim is evidence: the first day's lynch hits mafia in 13 of 23 games
against 4.3 expected ($3.0\times$, $p{=}4{\times}10^{-5}$). The lynched mafia are almost all non-claimers (the missing claim is
one cue among several: $\sim$40\% of the faction effect in a
conditional-logit decomposition). The edge is a property of the untrained model's prior and of the
protocol that exposes it. In the same chains' silent first games the day-1 lynch is at the
zero-information rate (7/25 vs 4.5, $p{=}0.18$); under turn-taking it
points below chance (1/23 vs 4.3, $p{=}0.10$): mafia who speak after claimers copy the claim
(2\% having heard none, 39\% after three), and the day's first
speaker, usually a citizen, draws the lynch
(Section~\ref{sec:behavior}). The 40 blank-notes redeployments replicate all three patterns on
fresh castings (broadcast 95\%/19\%, $3.2\times$; silent $p{=}1.0$;
sequential 4/40 vs 7.5).

\textbf{The collapse.} Societies un-learn the convention from the citizens' side, fast:
broadcast citizen claims fall $91\%\to21\%$ by generations 2--3 and
settle at 13\%, while mafia claims drop to $\le$0.6\% and never
recover --- the convention collapses at generation 2 in 21 of 23
broadcast chains, and first-day accuracy decays in step
($3.0\times\to1.7\times\to1.2\times$ chance). A late claim remains almost a certificate (mafia are 4 of 935 late
claimers), but with 87\% of citizens silent about their role,
non-claiming carries almost no information (lift $3.53\to1.12$): the
citizens' own change of convention removes the most informative
first-day cue.

\textbf{The notes carry the change, but not as doctrine.} Frozen
cross-deployment isolates the cause and prices it. On 40 identical
castings at $(16,3)$, blank-notes societies reproduce the untrained
pattern (95\% claims, first-day hits $3.2\times$) and score
$\hat\theta=+1.36$ \ci{+0.90}{+1.90} (town 31/40), while every
generation-60 note set, silence-evolved included, suppresses the
claims (2--24\%); five of six also remove the first-day edge (hits
$\le1.2\times$; the sixth, family C's broadcast notes, installs a
``framework'' cue, $3.73\times$), and every family scores below blank
(evolved $-$ blank $-1.37$ \ci{-1.97}{-0.83} for broadcast-trained,
$-1.25$ \ci{-1.85}{-0.73} for silence-trained; resolved for five of
six; family C $-0.51$ \ci{-1.24}{+0.13}), with the broadcast-over-silent deployment gap larger without notes
($+1.14$). Together with the
generation-1 result this locates the broadcast advantage in the
pretrained model: sixty generations of self-authored notes, at this
size and in redeployment, cost more than they add, and not only on
day 1 (days 2 and later: $-1.37$ \ci{-2.02}{-0.79}, $-1.02$
\ci{-1.68}{-0.45}). Yet the insight is \emph{written} early in only 9
of 56 chains (median first writing: generation 19), and generation-2
citizens whose notes say nothing about claims still fall to 18.8\%.
Every inherited note set we observed extinguished the naive
convention, whether or not it mentioned claims (a generic format
effect is not ruled out; notes that advise claiming push the other
way, 79\% claiming).

\textbf{The regime: no cycles; broadcast gives memory.} Across 75
chain and 30 cell series, no periodicity survives the AR(1)-surrogate null and correction:
the wave-like late curves of Fig.~\ref{fig:learning} are what
smoothing does to a persistent, aperiodic series. What is systematic is protocol-specific, statistic
by statistic. Broadcast hit ratios carry lag-1 memory (ACF $+0.14$,
$p{=}0.001$; $+0.17$ over silent, $+0.13$ over sequential; robust to
dropping the first ten generations and to excluding the
memory-affected games, but not replicated in the six broadcast chains
the defect never touched, which are also the smallest cells ---
exposure and size are confounded; supplement). All three
protocols fluctuate beyond a constant-edge noise floor, broadcast
resolvably more than silence ($+0.28$ log-variance). Casting-matched
families drift apart beyond a shared constant edge under broadcast in
generations 21--40 ($p{=}4.8{\times}10^{-4}$; over the full span the
protocol contrasts are borderline). Underneath, outcomes show no detected change of level (last third
vs middle third: 0 of 30 cells; early-vs-late tilt: 3 of 34 arms,
both directions, $\approx$1.7 expected by chance). The notes show no sign of converging in wording ($\approx$0.89
rewritten every generation, no downward trend; vocabulary overlap
rises, so this is surface change, not necessarily policy change; they
double in length under talk, 420$\to$897 words). One adapted outcome is a
deterioration: the $(16,5)$ broadcast collapse develops in the first
generations (generations 1--5: town 9 of 15, $\hat\theta$ $+1.17$
\ci{+0.34}{+2.02}; all generations: 26\%), as pre-registered;
mafia-role experience per lineage is five times higher at $m{=}5$
than at $m{=}1$, a confound this design cannot separate. Over sixty generations the notes keep changing while outcomes show
no systematic trend; what broadcast talk changes is the coupling:
generations cohere, families individuate.

\section{Behavior: Inside the Games}\label{sec:behavior}

What do half a million statements and 851k ballots show the societies
doing? Extractors are hand-validated blind (precision 50/50 on
claims; meme counts are lower bounds); everything here is an
association in the recorded games.

\textbf{Mafia rarely claim, and claimers fare worse.} Broadcast mafia claim to be
citizens in 60 of 27{,}972 statements (0.21\%, vs.\ 2.77\% for
citizens); the few who do are lynched \emph{more}, not less, the same
day (O/E 1.79 \ci{1.01}{2.68}). Under turn-taking mafia claim at
$0.46\times$ the citizen rate, and the claims are overwhelmingly copies rather than initiatives:
claim probability jumps from 0.005 to 0.41 (citizens) and 0.0003 to
0.24 (mafia) after an earlier claim, and 220 of 221 later-position
day-1 mafia lies follow one, consistent with documented LLM
conformity~\cite{zhu2025conformity}. Honest signals are costly too: broadcast
citizens who claim draw the night kill at O/E 1.37 \ci{1.21}{1.56}, and
the detective that announces itself is killed the following night 40 of
46 times (O/E 12.6) --- while no mafioso ever counterfeited a detective claim in 120
detective games, and confession is an endgame concession (87\% on the
final day).

\textbf{Turn-taking herds speech, not ballots.} Sequential statements cascade in the sense
of~\cite{bikhchandani1992cascades} (above), and the cascade carries
coordination without information: landslide days
($\ge$75\% of ballots on one seat) are as frequent as in broadcast
(13.8\% vs.\ 13.3\%) but land on mafia near chance ($1.08\times$ vs.\
$2.53\times$ for broadcast); the bandwagon index --- the probability that a ballot follows the
day's most-accused seat --- is indistinguishable (0.48 vs.\ 0.47,
$\approx$5$\times$ chance). When a sequential day opens with an accusation, later
\emph{statements} name the target more than in broadcast, yet
\emph{ballots} follow less (0.84 vs 0.62; 0.34 vs 0.43; lynched 0.38
vs 0.53). What
turn-taking anchors on is the first \emph{speaker}: named in 49\% of
later statements (broadcast 18\%), lynched at $1.66\times$
\ci{1.58}{1.73} the position-blind rate (broadcast placebo
$1.00\times$), mostly a citizen; first and last speakers are also
night-killed more often ($1.33\times$, $1.41\times$). Under turn-taking the channel carries exposure and imitation;
under broadcast, where statements are composed independently,
landslides are accurate, consistent with human evidence that exposure
to others' judgments erodes crowd accuracy~\cite{lorenz2011wisdom}.

\textbf{Night kills fall on the citizens who found the mafia.}
Citizens whose ballot went to a mafia member that day are
night-killed at risk ratios 1.10--3.26 (30 of 31 arms), most under
silence (9 of 9 cells), where no statements compete with the ballot
record. Ballot
accuracy tracks the protocol ordering of Section~\ref{sec:order}:
citizens' ballots land on mafia above chance in broadcast (9/9
cells), near chance in silence, and below chance in sequential play
(resolved in 4 of 9 cells). Mafia discipline completes the picture: partner ballots far below
chance (0.61\% vs 16.1\% at $k{=}0$), accusations at $0.30\times$
the citizen rate.

\textbf{The written culture.} The notes concentrate on
observables: vote-pattern tracking appears in 88.9\% of all 94k
notes, and
``analysts get killed'' --- the lesson of the night hazard above ---
rises from 27.8\% to 73.3\% of broadcast notes across the chains while peaking at 8.6\% in silent ones: the societies write down the dangers they can see.

\looseness=-1
\textbf{Take-away.} Claimants, accurate voters and the self-announced detective all
draw the night kill more often, consistent with the night removing
the honest signals of Section~\ref{sec:dynamics}. Turn-taking keeps coordination but
not information: its landslides account for about two-fifths of its
accuracy deficit, the rest coming from ordinary days ($0.82\times$ vs
$1.06\times$). These are associations, not tested mediators of
Section~\ref{sec:statics}'s contrasts.

\section{Related Work}\label{sec:related}

Mafia's theory: Braverman, Etesami and Mossel \cite{braverman2008}
prove the $\Theta(\sqrt R)$ fair line (with $R$ the population) for town play implemented by simultaneously announced random
numbers (in turn, they could be steered); our protocols vary the same
commit-before-observing structure for language
\cite{knight2026}; \cite{yao2008} gives recursion-based bounds and argues BEM's
multi-round conjecture conditionally (its information-zero exponent proved under weak-majority stopping
by \cite{tamashiro2026}, a partly machine-checked working paper);
\cite{migdal2010} gives the exact zero-information solution we use,
and \cite{stock2026} studies suspicion dynamics. The nearest analytic treatment of LLM Mafia play fits a capability
model to a four-player, single-day variant \cite{costa2026minimafia};
we anchor multi-day societies of 8--100 to Mafia's own exact
theory, with protocol, bandwidth and density as manipulanda. LLM social deduction more broadly optimizes agents
\cite{xu2024werewolf,xu2025latent,jin2024onuw}, improves frozen
agents through stored reflections \cite{xu2023communication},
evaluates at scale in fixed small games \cite{gamearena2026,
bailis2024werewolfarena}, times speech in asynchronous chat
\cite{eckhaus2025timetotalk}, and compares discussion against none
\cite{ogara2023hoodwinked}; in cooperative evaluation, one-by-one, simultaneous and summarized
exchange peak at similar accuracy \cite{chateval2024}; in our
competitive setting the format decides most of the value. Learning-to-communicate work trains the channel, mostly
cooperatively \cite{foerster2016learning,lowe2019pitfalls}, reaching
competition recently \cite{noukhovitch2021competition}; in RL Werewolf a learned signal helps one faction non-linearly
\cite{brandizzi2021rlupus}. Our contribution is not corpus
size but design: matched-casting protocol interventions
across sizes, measured against an exact reference, with an adaptation record
long enough to test dynamics against surrogate nulls.

\section{Discussion and Limitations}\label{sec:discussion}

\looseness=-1
\textbf{Reading.} The value of communication here is a property of
the \emph{protocol}: simultaneous composition keeps the channel
informative; under turn-taking it carries imitation and exposure;
under silence the minority's ballots and night choices
dominate. What adaptation adds is protocol-specific, transfers unstably, and
shows no systematic improvement: its fastest, most uniform act removes the most informative
first-day cue, written culture trailing it; regime-wise, no detected
trend under continual rewriting, family individuation only under
broadcast. For multi-agent LLM design: more communication is not always better (far less when speakers
react first), and accumulated self-generated experience can
subtract value while appearing to encode lessons
\cite{xiong2026memory}.

\looseness=-1
\textbf{Limitations.} One model at pinned settings (one injection
probe); the generation-1 convention and its collapse are properties of
this model's prior and prompt stack (no transfer claim); self-written notes are one adaptation path.
Frozen bandwidth curves use one family per size amid large
heterogeneity; the pre-registered matrix added two families and refuted the
single-family cross-over rule (\S\ref{sec:transfer}). Each $(70,6)$ arm is one family; the $(46,5)$ silent comparator is
non-concurrent; the scalar tilt under-predicts the talking arms'
wins there (\S\ref{sec:scale}). Dynamics tests use one game per generation per chain (strong
periodicity excluded, not every cycle); behavior results are
associations under selection, extractor recall bounded and
disclosed. The planned interventions on nights, ballots and history were not
run. The nine headline contrasts survive Holm and BH (Bonferroni
marginal at $(8,2)$); broadcast$-$silent is positive in 23 of 23
families, resolved in 18 (supplement).

\section{Conclusion}\label{sec:conclusion}

\looseness=-1
Against Mafia's exact zero-information reference, these LLM
societies profit from statements composed independently and
simultaneously; reaction-first discussion carries imitation and
exposure instead; silence favors the informed minority; and limited
talk is worth less at 70 players than at 46, below silence at $k{=}8$. Read in time, they abandon their most
informative convention within two generations and keep rewriting
without converging --- the protocol setting what a collective can
infer and how that changes, while open talk's cost keeps growing
after its value, at the bandwidths that fit, is gone.

\balance
\bibliographystyle{ACM-Reference-Format}
\bibliography{refs}

\end{document}